\documentclass[float,epsfig,usenatbib,a4]{mn2e}
\usepackage{graphicx}
\usepackage{appendix}
\usepackage{psfrag}
\usepackage{amsmath,amssymb}
\usepackage{threeparttable}
\usepackage{float}
\usepackage{subfigure}
\usepackage{longtable}
\usepackage{lscape}
\usepackage{afterpage}
\usepackage{url}

\title[PACS photometry of the {\it Herschel} Reference Survey]{PACS photometry of the {\it Herschel} Reference Survey -- Far-infrared/sub-millimeter 
colours as tracers of dust properties in nearby galaxies\thanks{{\it Herschel} is an ESA space observatory with science instruments provided by 
European-led Principal Investigator consortia and with important participation from NASA.}}
\author[L. Cortese et al.]
{L. Cortese\thanks{lcortese@swin.edu.au}$^{1,2}$, J. Fritz$^3$, S. Bianchi$^4$, A. Boselli$^5$, L. Ciesla$^6$, G.~J. Bendo$^7$, M. Boquien$^{5,8}$, \newauthor
H. Roussel$^9$, M. Baes$^3$, V. Buat$^5$, M. Clemens$^{10}$, A. Cooray$^{11}$, D. Cormier$^{12}$, \newauthor  J.~I. Davies$^{13}$, 
I. De Looze$^3$, S.~A. Eales$^{13}$, C. Fuller$^{13}$, L.~K. Hunt$^4$,  S. Madden$^{14}$,\newauthor J. Munoz-Mateos$^{15}$, C. Pappalardo$^4$, 
D. Pierini$^{16}$, A. R\'emy-Ruyer$^{14}$, M. Sauvage$^{14}$,\newauthor  S. di Serego Alighieri$^4$, M.~W.~L. Smith$^{13}$, L. Spinoglio$^{17}$, M. Vaccari$^{18}$, C. Vlahakis$^{19}$ \\
$^1$Centre for Astrophysics \& Supercomputing, Swinburne University of Technology, Mail H30 - PO Box 218, Hawthorn, VIC 3122, Australia\\
$^2$European Southern Observatory, Karl-Schwarzschild Str. 2, 85748 Garching bei Muenchen, Germany\\
$^3$Sterrenkundig Observatorium, Universiteit Gent, Krijgslaan 281 S9, 9000, Gent, Belgium\\
$^4$INAF-Osservatorio Astrofisico di Arcetri, Largo Enrico Fermi 5, 50125 Firenze, Italy \\
$^5$Laboratoire d'Astrophysique de Marseille - LAM, Universit\'e d'Aix-Marseille \& CNRS, UMR7326, \\ 38 rue F. Joliot-Curie, F-13388 Marseiulle Cedex 13, France \\
$^6$University of Crete, Department of Physics, Heraklion, Crete, 71003, Greece\\
$^7$UK ALMA Regional Centre Node, Jodrell Bank Centre for Astrophysics, School of Physics and Astronomy, University of Manchester, \\ Oxford Road, Manchester M13 9PL, United Kingdom \\
$^8$Institute of Astronomy, University of Cambridge, Madingley Road, Cambridge CB30HA, UK\\
$^9$Institut d'Astrophysique de Paris, Universit{\'e} Pierre et Marie Curie (UPMC), CNRS (UMR7095), 75014 Paris, France\\
$^{10}$Osservatorio Astronomico di Padova, Vicolo dell'Osservatorio 5, I-35122 Padova, Italy\\
$^{11}$University of California, Irvine, Department of Physics \&  Astronomy, 4186 Frederick Reines Hall, Irvine, CA, USA\\
$^{12}$Institut f\"ur Theoretische Astrophysik, Zentrum f\"ur Astronomie der Universit\"at Heidelberg, Albert-Ueberle Str. 2, D-69120 Heidelberg, Germany\\
$^{13}$School of Physics and Astronomy, Cardiff University, The Parade, Cardiff, CF24 3AA, UK \\
$^{14}$Laboratoire AIM, CEA, Universit\'{e} Paris VII, IRFU/Service d'Astrophysique, Bat. 709, 91191 Gif-sur-Yvette, France\\
$^{15}$European Southern Observatory, Alonso de Cordova 3107, Vitacura, Casilla 19001, Santiago de Chile\\
$^{16}$Max-Planck-Institut f\"ur extraterrestrische Physik, Giessenbachstrasse, Postfach 1312, D-85741, Garching bei M\"unchen, Germany\\
$^{17}$Istituto di Fisica dello Spazio Interplanetario, INAF, Via Fosso del Cavaliere 100, I-00133 Roma, Italy\\
$^{18}$ Astrophysics Group, Physics Department, University of the Western Cape, Private Bag X17, Bellville 7535, Cape Town, South Africa\\
$^{19}$Joint ALMA Observatory/European Southern Observatory, Alonso de Cordova 3107, Vitacura, Santiago, Chile\\
}

\date{Accepted 2014 January 21.  Received 2014 January 20; in original form 2013 November 3}

\begin{document}
\newcommand{\Zsolar}{\mbox{$\,\rm Z_{\odot}$}}
\newcommand{\Msolar}{\mbox{$\,\rm M_{\odot}$}}
\newcommand{\Lsolar}{\mbox{$\,\rm L_{\odot}$}}
\newcommand{\xs}{$\chi^{2}$}
\newcommand{\dxs}{$\Delta\chi^{2}$}
\newcommand{\xsn}{$\chi^{2}_{\nu}$}
\newcommand{\ls}{{\tiny \( \stackrel{<}{\sim}\)}}
\newcommand{\gs}{{\tiny \( \stackrel{>}{\sim}\)}}
\newcommand{\asec}{$^{\prime\prime}$}
\newcommand{\amin}{$^{\prime}$}
\newcommand{\mstar}{\mbox{$M_{*}$}}
\newcommand{\hi}{H{\sc i}}
\newcommand{\hii}{H{\sc ii}\ }
\newcommand{\kms}{km~s$^{-1}$\ }

\maketitle

\label{firstpage}

\begin{abstract}
We present {\it Herschel}/PACS 100 and 160 $\mu$m integrated photometry for the 323 galaxies in the {\it Herschel} Reference Survey (HRS), a 
K-band-, volume-limited sample of galaxies in the local Universe. Once combined with the {\it Herschel}/SPIRE observations already available, 
these data make the HRS the largest representative sample of nearby galaxies with homogeneous coverage across the 100-500 $\mu$m 
wavelength range. In this paper, we take advantage of this unique dataset to investigate the properties and shape of the far-infrared/sub-millimeter 
spectral energy distribution in nearby galaxies. 
We show that, in the stellar mass range covered by the HRS (8\ls$\log(M_{*}/M_{\odot})$\ls12), the far-infrared/sub-millimeter 
colours are inconsistent with a single modified black-body having the same dust emissivity index $\beta$ for all galaxies. 
In particular, either $\beta$ decreases, or multiple temperature components are needed, when moving from metal-rich/gas-poor 
to metal-poor/gas-rich galaxies. We thus investigate how the dust temperature and mass obtained from a single modified black-body 
depend on the assumptions made on $\beta$. We show that, while the correlations between dust temperature, 
galaxy structure and star formation rate are strongly model dependent, the dust mass scaling relations are 
much more reliable, and variations of $\beta$ only change the strength of the observed trends.
\end{abstract}
\clearpage
\begin{keywords}
galaxies: fundamental parameters -- galaxies: ISM -- infrared: galaxies 
\end{keywords}

\section{Introduction}
It is now well established that approximately 
half of the radiative energy produced by galaxies is 
absorbed by dust grains and re-emitted in the infrared regime \citep{hauser01,bosellised,dole06,dale07,burga2013}. 
Thus, observations in the $\sim$10-1000 $\mu$m wavelength range 
provide us with a unique opportunity not only to quantify half of the bolometric 
luminosity of galaxies, but also to characterise the properties of cosmic dust.
Moreover, since dust grains are crucial for the star formation cycle \citep{hollenbach71}, such information 
can give us important insights into the physical 
processes regulating galaxy evolution (e.g., \citealp{dunne11}). 

Unfortunately, despite its paramount importance, we are still missing a complete 
and coherent picture of dust properties in galaxies 
across the Hubble sequence, and of the exact role played 
by grains in regulating star formation \citep{mckee10}. 
Indeed, we know very little about the dust composition 
in galaxies outside our own Local Group \citep{draine07b,compiegne11} and if/how 
it is regulated by the physical conditions experienced by 
grains in the inter-stellar medium (ISM). 
Hence, our estimates of dust masses in galaxies are still 
highly uncertain \citep{fink99,dupac03,gordon10,paradis10,plankbeta}. 

Luckily, the last decade has seen the start of a golden age for 
observational far-infrared (FIR) and sub-millimeter (submm) astronomy, providing a new boost 
to the refinement of theoretical dust models \citep{meny07,draine07b,hoang2010,compiegne11,steinacker13}. 
In particular, the {\it Spitzer} \citep{spitzer}, and more recently 
{\it Herschel} \citep{pilbratt10} and {\it Planck} \citep{planck} space telescopes 
are finally gathering a wealth of information on the dust emission 
from thousands of galaxies up to $z\sim$2. 
Particularly important for a proper characterisation of 
dust in galaxies is the radiation emitted at wavelengths  
\gs100-200 $\mu$m. In this regime, the integrated emission from galaxies 
originates predominantly from dust in thermal equilibrium, heated by the diffuse interstellar 
radiation field (ISRF), which represents the bulk of the dust mass 
in a galaxy (e.g., \citealp{sodrosky89,sauvage92,calzetti95,walterbos,bendo10,boquien11,bendo12b}). Thus, by characterising the dust emission in the \gs100 $\mu$m 
wavelength domain, we have a unique opportunity to provide strong 
constraints to theoretical models, and to refine our census of the 
dust budget in galaxies. 

The first natural step in this direction is to quantify how 
the shape of the dust spectral energy distribution (SED) varies 
with galaxy properties across a wide range of morphological type, 
star formation activity, cold gas mass and metal content. 
This is necessary to determine if the amount 
of radiation emitted at each wavelength is simply regulated by 
the intensity of the ISRF responsible for the dust heating, or 
whether it retains an imprint of the chemical composition 
of the grains. Indeed, only after a careful characterisation of 
the physical parameters regulating the dust SED, will it be possible 
to properly convert observables into physical quantities such 
as dust temperatures and dust masses. 

Many recent works \citep{gordon10,skibba11,davies11,plankbeta,galametz12,auld13} have shown 
that, above $\sim$100 $\mu$m, the dust SED is very well approximated by a simple modified black-body (but see also \citealp{bendo12b}):
\begin{equation}
F_{\nu} = \frac{M_{dust}}{D^{2}} \kappa_{\nu_{0}}\left(\frac{\nu}{\nu_{0}}\right)^{\beta} B_{\nu}(T)
\label{eq1}
\end{equation}
where $F_{\nu}$ is the flux density emitted at the frequency $\nu$, $\kappa_{\nu_{0}}$ is the dust mass absorption 
coefficient at the frequency $\nu_{0}$, $\beta$ gives its variation as a function of frequency, 
$D$ is the galaxy distance and $B_{\nu}(T)$ is the Planck function.
Mounting evidence is emerging that $\beta$ is not the same in all galaxies (e.g., \citealp{remy13}), 
and may also vary within galaxies (e.g., \citealp{galametz12,smith12b}). 

Modified black-bodies are simple models and cannot 
properly reproduce real dust properties (e.g., \citealp{draine07b,shetty09b,bernard10}). 
Several dust components at various temperatures contribute to the total emission along the lines-of-sight. 
This implies the presence of temperature mixing that can cause variations
of the infrared slope, and thus in the apparent emissivity index $\beta$.
Nevertheless, parameterization of the dust SEDs through modified black-body 
fitting is a powerful tool to help understand variations of dust properties 
with other galaxy characteristics, especially in case of sparse sampling of 
the FIR/sub-mm wavelength range (e.g., high-redshift galaxies \citealp{magdis2011,symenodis2013}).
Therefore, it is extremely important to determine in which cases a single modified back-body 
can be used, and how temperature and dust mass estimates are affected by 
the assumptions made on $\beta$.

In order to ascertain the dust properties of galaxies in the local Universe, and 
to provide new constraints to theoretical models, we have carried out the {\it Herschel} Reference Survey (HRS, \citealp{HRS}), 
a {\it Herschel} guaranteed time project focused on the study of the interplay between dust, gas and star formation 
in a statistically significant sample of $\sim$300 galaxies spanning a wide range of morphologies, stellar masses (8\ls log(M$_{*}$/M$_{\odot})$\ls 12), 
cold gas contents (-3\ls log(M$_{HI}$/M$_{*}$)\ls 1), metallicities (8.2\ls 12+log(O/H) \ls 8.9), 
and specific star formation rates (-12 \ls log(SFR/M$_{*}$)\ls -9). 
The combination of {\it Herschel}/SPIRE \citep{spire} observations with the multi-wavelength dataset we have been 
assembling \citep{ciesla12,hrsgalex,boselli13,hughes13}, 
has already allowed us to have a first glimpse at how the dust content and shape of the dust SED vary with 
internal galaxy properties \citep{boselli10,boselli12,cortese12}. 
In particular, \cite{boselli10,boselli12} have shown that the slope of the dust SED 
in the 200-500 $\mu$m interval decreases from $\beta\sim$2 to $\beta\sim$1 when moving 
from metal-rich to metal-poor galaxies. However, our analyses have so far been limited by the lack of 
data in the $\sim$100-200 $\mu$m wavelength range for the entire sample.  

Thus, in this paper we present integrated {\it Herschel}/PACS \citep{pacs} 100 and 160 
$\mu$m flux densities for all the HRS sample and take advantage of our multiwavelength 
dataset to perform a first analysis 
of the properties of the dust SED across our entire sample.
Corresponding to the peak of the dust SED, the 100-200 $\mu$m wavelength interval 
is crucial not only to properly quantify the shape of the SED, but also 
to accurately determine the average dust temperature and total dust mass in galaxies.
 These data make 
the HRS the largest representative sample of nearby galaxies with homogeneous 
coverage across the $\sim$100-500 $\mu$m wavelength range. 
In addition to releasing our dataset to the community, our primary goals are 1) to 
investigate how the shape of the dust SED varies with internal galaxy properties, and 2) to determine 
whether the integrated dust SED of HRS galaxies can always be reduced to a single modified 
black-body with a constant value of $\beta$ and, if not, what are the possible biases introduced 
by this assumption. The results of SED fitting with the dust models of \cite{draine07} will 
be presented in a forthcoming paper (Ciesla et al., submitted.).

This paper is organized as follows. In Sect. 2 we describe the {\it Herschel} observations, data reduction, flux density estimates and comparison with the literature. 
In Sect. 3 we use the PACS and SPIRE colours to investigate how the shape of the dust SED varies with internal galaxy properties. 
In Sec. 4, we show how the dust temperature and mass obtained from fitting a single modified black-body to the {\it Herschel} data depend 
on the assumptions made on $\beta$. Finally, the summary and implications of our results are presented in Sec. 5.

\begin{figure*}
\centering
\includegraphics[width=17.5cm, angle=0]{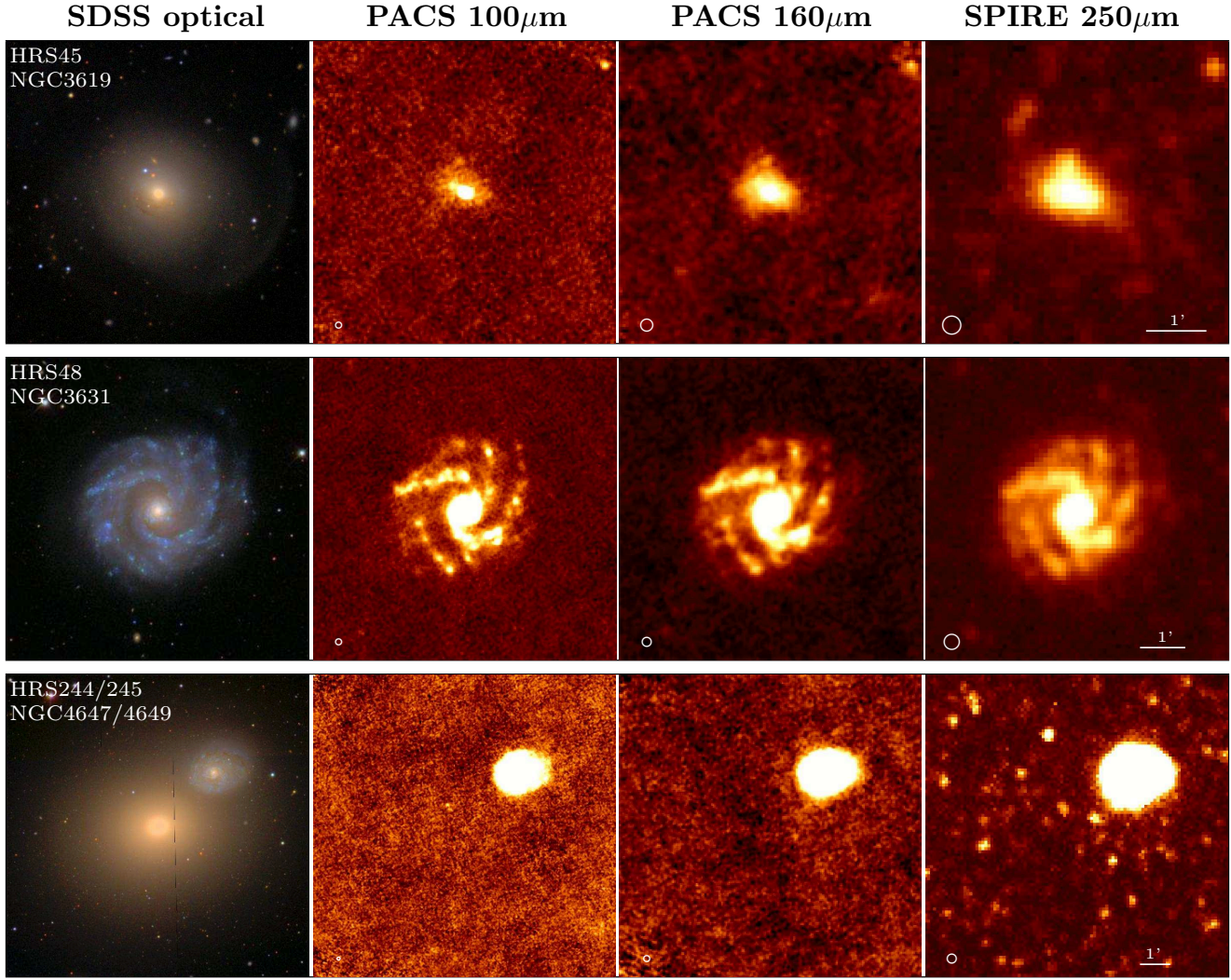}
\caption{Comparison of the quality of our PACS images with the Sloan Digital Sky 
Survey optical and SPIRE 250 $\mu$m images. We show three types of objects: an early-type with dust lanes (top row), an unperturbed 
late-type spiral and an un-detected elliptical and its spiral companion. The size of the PACS and SPIRE beams 
is shown in the bottom left corner of each panel.}
\label{images}
\end{figure*}

\section{The data}
\subsection{The {\it Herschel} Reference Survey}
The HRS is a volume-limited sample (i.e., 15$\leq$$D$$\leq$25 Mpc) including all late-type galaxies (261 Sa and later) 
with  2MASS \citep{2massall} K-band magnitude K$_{S,tot} \le$ 12 mag and all early-type galaxies (62 S0a and earlier) 
with K$_{S,tot} \le$ 8.7 mag\footnote{We note that one galaxy (HRS228) had a wrong redshift reported in NED, and is in reality a background galaxy. In this work, 
we have included it for completeness.}. Additional selection criteria are high galactic latitude ($b>$ +55$^{\circ}$) 
and low Galactic extinction ($A_{B}$ $<$ 0.2 mag, \citealp{schlegel98}), to minimize Galactic cirrus contamination. 
More details on the original selection can be found in \cite{HRS}, while the most recent morphological classifications 
and distance estimates are presented in \cite{hrsgalex}.

\subsection{PACS observations and data reduction}
The {\it Herschel}/PACS 100 and 160 $\mu$m observations of HRS galaxies presented in this work have been 
obtained as part of various open-time {\it Herschel} projects.

The vast majority of the data (228 out of 323 galaxies) comes from our own {\it Herschel} cycle 1 open time proposal (OT1\_lcortese1).  
Each galaxy was observed in scan mode, along two perpendicular axes, at the medium scan speed of 20\arcsec/sec.
Two repetitions were done in each scan direction. The size of each map was chosen to match the size 
of our SPIRE images (see \citealp{ciesla12}), making sure to have homogeneous coverage across the entire 100-500 $\mu$m range. 

Maps for additional 83 HRS galaxies have been obtained as part of the {\it Herschel} Virgo Cluster Survey (HeViCS, \citealp{davies10}).
HeViCS mapped the Virgo cluster with both PACS and SPIRE simultaneously at the fast scan speed of 60\arcsec/sec.
The observing strategy consists of scanning each $\sim$4$\times$4 deg$^{2}$ field in two orthogonal directions, and repeating each scan 
four times \citep{auld13}. The faster scan speed of the {\it Herschel} parallel mode with respect to 
the scan map mode, used for our observations, is compensated by the higher number of repetitions performed in the Virgo cluster, 
making the two datasets highly comparable (i.e., within $\sim$30\%) in terms of their final noise. 

PACS observations for the remaining 12 HRS galaxies have been retrieved from the {\it Herschel} public archive, and 
come from various projects (i.e., \citealp{kingfish}, KPGT\_esturm\_1, OT1\_acrocker\_1, OT2\_emurph01\_3, GT1\_lspinogl\_2, OT2\_aalonsoh\_2). 
All data have been obtained in scan mode at the medium scan speed of 20\arcsec/sec and they reach a noise level similar or lower than 
our own observations. For one galaxy (HRS3) only 160 $\mu$m observations 
are available as the object lies at the edge of the 100 $\mu$m map, making the data not suitable for accurate photometry.  
Thus, in summary, all 323 galaxies in the HRS have been observed at 160 $\mu$m, whereas 100 $\mu$m data are available for 322 objects. 

All raw PACS data were processed from Level-0 to Level-1 within HIPE (v10.0.0, \citealp{hipe}) using the calibration file v48. 
This pre-processing includes, among the other tasks, pixel flagging, flux density conversion and coordinate assignment. 
To remove the 1/$f$ noise which, at this point, still dominates the timelines, the Level-1 data were fed into \texttt{Scanamorphos} (version 21, \citealp{scanam}), 
an IDL algorithm which performs an optimal correction by exploiting the redundancy in the observations of each sky pixel. 
No noise modelling is hence needed.
The pixel size of the final maps was chosen to sample at the best the point-spread-function, at the respective wavelengths, typical of the 
data taken at medium scan speed: 1.7 and 2.85 arcsec pixel$^{-1}$ at 100 and 160 $\mu$m, respectively (i.e., FWHM/4). 
The typical pixel-by-pixel noise in the map varies between $\sim$0.1 and $\sim$0.25 mJy pixel$^{-1}$ at 160 $\mu$m 
and between $\sim$0.04 and $\sim$0.1 mJy pixel$^{-1}$ at 100 $\mu$m. 

In order to show the data quality of the new observations presented here, in Fig.~\ref{images} we 
compare the PACS images for three of our targets with the RGB Sloan Digital Sky Survey \citep{sdssDR7} optical and SPIRE 250 $\mu$m \citep{ciesla12} images. 
We show an example of an early-type galaxy with dust lanes (HRS45, top row), late-type galaxy (HRS48, middle row) and 
un-detected elliptical and its spiral companion (HRS244/245, bottom row).

\subsection{PACS 100 and 160 $\mu$m integrated photometry}
Integrated 100 and 160 $\mu$m photometry has been performed following very closely the technique 
used by \cite{ciesla12} for the SPIRE data of HRS galaxies. This is crucial to properly 
combine the two datasets, and to characterise the shape of the SED across the entire 100-500 $\mu$m wavelength range. 
Thus, whenever possible, we determined integrated flux densities within the same apertures 
adopted in \cite{ciesla12}. The aperture sizes are adapted to include the entire extent of the FIR emission 
from the galaxies, and they correspond to $\sim$1.4, $\sim$0.7 and $\sim$0.3 
times the optical diameter for late-type, lenticular and elliptical galaxies, respectively. 
Only for 36 galaxies ($\sim$11\% of the sample) we choose different sizes 
than those used for SPIRE. There are three different reasons why we did so: 
a) For 23 galaxies (HRS6, 14, 22, 32, 67, 71, 75, 158, 209, 223, 225, 238, 243, 249, 255, 257, 261, 264, 286, 300, 315, 317, 322) 
the 100 and 160 $\mu$m emission is significantly less extended than the size of the aperture used by \cite{ciesla12}. 
Although this does not affect the estimate of the integrated flux density, it artificially boosts the error associated 
with our measurements to values always above 50\%, and sometimes even higher than 100\%. 
Thus, for these objects, we reduced the size of the aperture (on average by $\sim$26\%) to obtain more realistic error estimates. 
We note that the size chosen is still larger than the extent of the FIR emission (so 
that aperture corrections are not necessary), and that the flux density estimated within these new apertures 
is consistent with the value obtained using \cite{ciesla12} apertures. 
b) 10 galaxies (HRS7, 68, 129, 138, 161, 174, 210, 231, 258, 308) were not spatially 
resolved in the SPIRE bands, and SPIRE photometry was carried out directly on the time-line data. 
For these cases, which are generally resolved by PACS, we chose new apertures which include all the emission from the target. 
c) For 3 galaxies (HRS4, 122, 263), the PACS maps available from the archive were slightly smaller than our 
SPIRE maps. While these maps are large enough to include the entire aperture used in \cite{ciesla12}, no 
space is left to properly estimate the background. Thus, the aperture has been reduced in order to allow a more accurate 
background estimate, and still encompass all the emission from the galaxy. 

Sky background was determined in fifteen to thirty regions, depending on the size of the target, around the chosen aperture. The use of 
various regions instead of just a circular annulus makes it easier to estimate the 
large scale variations in the background and to avoid background/foreground sources around the target. 
The mean sky value was then subtracted from each map before performing the flux density extraction. 
Since cirrus contamination is significantly less of an issue than in SPIRE images, we did 
not find necessary to perform a more complex modelling of the background. However, as discussed below, 
the effect of any residual large scale gradient is included in our error estimates. 

Errors on integrated flux densities have been estimated following the guidelines described 
in \cite{scanam}, which are consistent with what is done in \cite{ciesla12} for HRS SPIRE data. 
Briefly, there are three sources of errors that affect our measurements:
\begin{equation}
\sigma_{tot} = \sqrt[]{\sigma_{cal}^{2} + \sigma_{instr}^{2} + \sigma_{sky}^{2}} 
\end{equation}
where $\sigma_{cal}$ is the flux calibration uncertainty (here assumed to be
5\%; \citealp{pacs_calibration}), $\sigma_{instr}$ 
is the instrumental noise which depends on the number of scans crossing a pixel, and is obtained 
by summing in quadrature the values on the error map within the chosen aperture, and $\sigma_{sky}$ 
is the error on the sky measurement. As discussed in \cite{scanam}, the sky uncertainty 
results from the combination of the uncorrelated error on the mean value of the sky ($\sigma_{skypix}$ i.e., the 
pixel-to-pixel variation across the image), and the correlated noise due to long time-scale 
drift residuals responsible for the large scale structures present in the image background 
($\sigma_{skymean}$ i.e., the standard deviation of the mean value of the sky measured in different apertures around the galaxy; 
see also \citealp{bosiso03,armando05}).
In detail, 
\begin{equation}
\sigma_{sky} = \sqrt[]{N_{ap}\sigma_{skypix}^{2} + N_{ap}^{2}\sigma_{skymean}^{2}} 
\end{equation}
where $N_{ap}$ is the number of pixels in the aperture used to integrate the galaxy flux density. 
As expected, for the vast majority of our objects the dominant source of error is the correlated 
uncertainty on the large-scale structure of the background. 
The average total uncertainties are $\sigma_{tot}$$\sim$16\% and 12\% at 100 and 160 $\mu$m, respectively. 

Out of the 323 galaxies observed, 282 have been detected in both bands (284 at 160 $\mu$m only).
This matches the HRS detection fraction in the SPIRE bands (i.e., 284 galaxies detected 
at 250 $\mu$m), allowing us to characterise the shape of the FIR/sub-mm SED across the entire 100-500 $\mu$m range for 
almost 300 galaxies. In case of non detections, upper limits have been estimated as 
3$\times\sigma_{tot}$, using the same apertures as in \cite{ciesla12}.  
\begin{figure*}
\centering
\includegraphics[width=17.8cm]{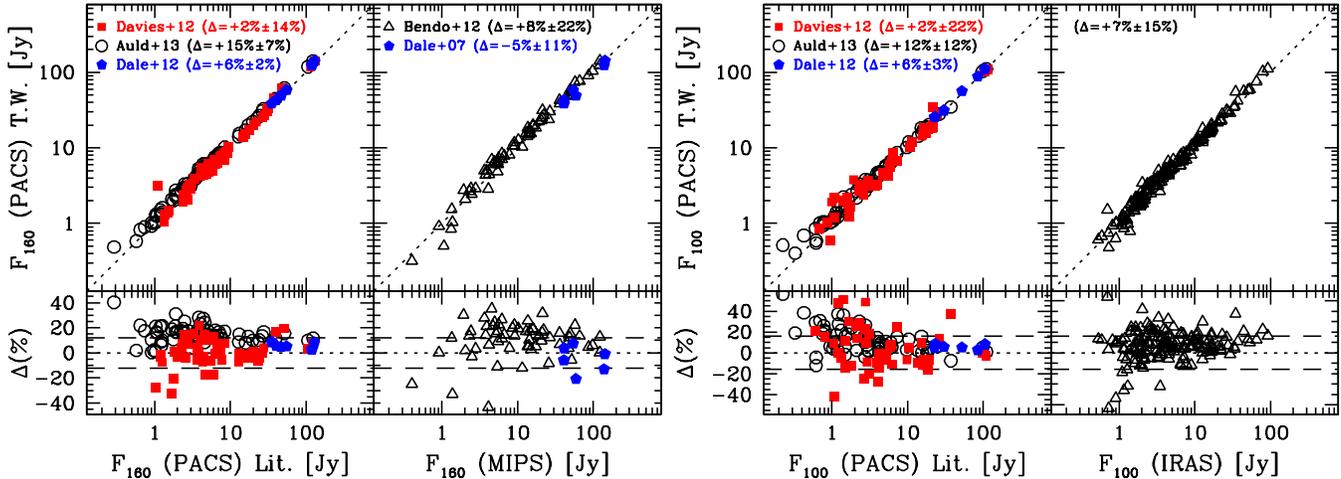}
\caption{Comparison between our 160 $\mu$m (left) and 100 $\mu$m (right) flux density estimates and those presented in the literature. 
The bottom panels show the difference (this work (T.W.) -literature) in percentage for each dataset. 
For each PACS channel, the left panel shows the comparison with literature estimates based 
on PACS data, while in the right panel the comparison with {\it Spitzer}/MIPS and IRAS observations is presented. 
The dotted lines indicate the one-to-one relation, and the dashed lines the average uncertainty in our 
flux density estimates.}
\label{litcomp}
\end{figure*}

The results of our photometry are presented in Table~\ref{tab}. 
The columns are as follows: 

Columns 1-6:  HRS \citep{HRS}, CGCG \citep{ZWHE61}, VCC \citep{vcc}, UGC \citep{ugc}, NGC \citep{ngc} and IC \citep{ic} names. 

Columns 7-8: the J2000 right ascension and declination.

Column 9: Morphological type, taken from \cite{hrsgalex}: -2=dE/dS0, 0=E-E/S0, 1=S0, 2=S0a-S0/Sa, 3=Sa, 4=Sab, 5=Sb, 6=Sbc, 7=Sc, 8=Scd, 
9=Sd, 10=Sdm-Sd/Sm, 11=Sm, 12=Im, 13=Pec, 14=S/BCD, 15=Sm/BCD, 16=Im/BCD, 17=BCD.

Column 10: 100 $\mu$m flux density measurement flag. Non detections=0, Detections=1, Confused (i.e., flux density estimate 
significantly contaminated by the presence of another object)=2. For confused galaxies, flux densities should 
be considered as an upper limit to the real value.

Column 11: Integrated 100 $\mu$m flux density, or upper limit in Jy.

Column 12: Total uncertainty on the 100 $\mu$m flux density measurement in Jy.

Column 13: 160 $\mu$m flux density measurement flag. 

Column 14: Integrated 160 $\mu$m flux density, or upper limit in Jy.

Column 15: Total uncertainty on the 160 $\mu$m flux density measurement in Jy.

Columns 16-18: Major, minor semi-axis  (in arcseconds) and position angle (in degrees) of the aperture used for the photometry.

Column 19: {\it Herschel} Proposal ID. 

This table, as well as all the reduced PACS maps, are publicly available on the Herschel Database in Marseille (HeDaM, \url{http://hedam.oamp.fr/}).  

\subsection{Comparison with the literature}
In order to check the reliability of the PACS flux density measurements presented here, we compare our far-infrared 
integrated flux densities with the values presented in the literature, which are based 
on PACS, {\it Spitzer}/MIPS or IRAS observations.
The results of these comparisons are shown in Fig.~\ref{litcomp}.

The difference between our flux density estimates and those presented in \cite{dale12} is $\sim$+6\% (standard deviation of $\sim$2-3\%), with our flux densities being 
brighter, although the number statistics is very small (6 galaxies in total). 
This difference is within the quoted uncertainties, and is mainly due to the different technique 
used to estimate flux densities (i.e., different background apertures and the use of aperture corrections 
not adopted in this work).

\cite{auld13} recently published PACS flux density measurements for all the VCC galaxies in 
the HeViCS footprint. A comparison between the flux density estimates for the 65 detected galaxies in common reveals a nice correlation 
between the two estimates with a standard deviation of just $\sim$12\% and $\sim$7\% at 
100 and 160 $\mu$m, respectively. However, \cite{auld13} measurements are systematically 
$\sim$12\% and $\sim$15\% lower than ours. 

After various tests, we concluded that there are two main reasons for this discrepancy. 
First, a different flux density estimate technique. \cite{auld13} used apertures on average significantly smaller than ours (e.g., see 
their Fig.~3), and then applied aperture corrections. Indeed, by using our own apertures on the \cite{auld13} dataset, 
we find no systematic offset with our 100 $\mu$m data, whereas at 160 $\mu$m there is still a difference of $\sim$12\%. 

Second, a different data reduction technique. \cite{auld13} used the {\it naive} projection task \texttt{photProject} in HIPE to reduce PACS images. 
This requires the use of a high-pass filter to correct for 1/$f$ noise, and such procedure could remove diffuse emission associated to extended objects. 
By using the same apertures on the HeViCS maps reduced 
with both \texttt{photProject} and \texttt{Scanamorphos}, we find that \texttt{photProject} maps provide flux densities $\sim$10\% 
lower than those obtained with \texttt{Scanamorphos}, while no difference is seen at 100 $\mu$m.
Thus, the remaining difference at 160 $\mu$m is due to the use of \texttt{photProject} instead of \texttt{Scanamorphos}. 
Indeed,  as mentioned above, this is likely due to the use of high-pass filtering which removes diffuse emission, much more 
commonly present at 160 $\mu$m than at 100 $\mu$m (see also \citealp{remy13}).

We also compared our measurements to those presented by \cite{davies11} for the 49 galaxies in common. 
These are based on an early HeViCS data release and are measured on apertures much more similar to the ones we used. 
Our flux density measurements agree very 
well with these estimates ($\sim$+2$\pm$22\% and $\sim$+2$\pm$14\% at 100 and 160 $\mu$m, respectively). 
The scatter is larger than in the case of \cite{auld13}, but consistent with the typical uncertainty given 
in \cite{davies11}. It is likely that, in this case, the different calibration between the two datasets 
compensates for the intrinsic differences between \texttt{photProject} and \texttt{Scanamorphos}, 
providing a set of measurements consistent with our own.

{\it Spitzer}/MIPS 160 $\mu$m flux densities for 103 galaxies in the HRS have been published by 
\cite{bendo12}. In order to perform a proper comparison with our data, we removed those 
galaxies which were flagged as problematic due to incomplete coverage, 
or simply being confused with other nearby galaxies of similar surface brightness in \cite{bendo12}. For the remaining 65 objects in common 
our flux densities are $\sim$8\% brighter than those of MIPS one, with quite a large scatter ($\sim$22\%). 
This large scatter is mainly due to two galaxies (which fall outside the residual plot in Fig.~\ref{litcomp}): HRS129, 258. A comparison between the PACS, SPIRE and MIPS data for these 
galaxies shows that the MIPS data suffer from background confusion effects, making it difficult to
separate emission from the target and background sources. Moreover, the MIPS 
observations for these galaxies were performed in photometry mode, 
which produces compact maps where it is difficult to measure the background.
Once these are removed from the sample, the difference between MIPS and PACS measurements becomes $\sim$+10$\pm$14\%. 
Conversely, the comparison with the {\it Spitzer}/MIPS 160 $\mu$m flux densities 
presented in \cite{dale07} for the 6 SINGS galaxies in our sample shows an average difference 
of $\sim$-5$\pm$11\%. All these values are within the 12\% flux calibration uncertainty in MIPS data \citep{spitzercal}. 
A PACS-to-MIPS 160 $\mu$m flux density ratio systematically higher 
than 1 has also been found by comparing pixel-by-pixel photometry of nearby galaxies \citep{aniano12,draine13}. 

We can thus conclude that our 160 $\mu$m PACS flux density measurements 
are consistent with those of {\it Spitzer}/MIPS within $\sim$20\%, in agreement with the results 
obtained by the PACS Team \citep{pacs_mips}.
\begin{figure*}
\centering
\includegraphics[width=17.8cm]{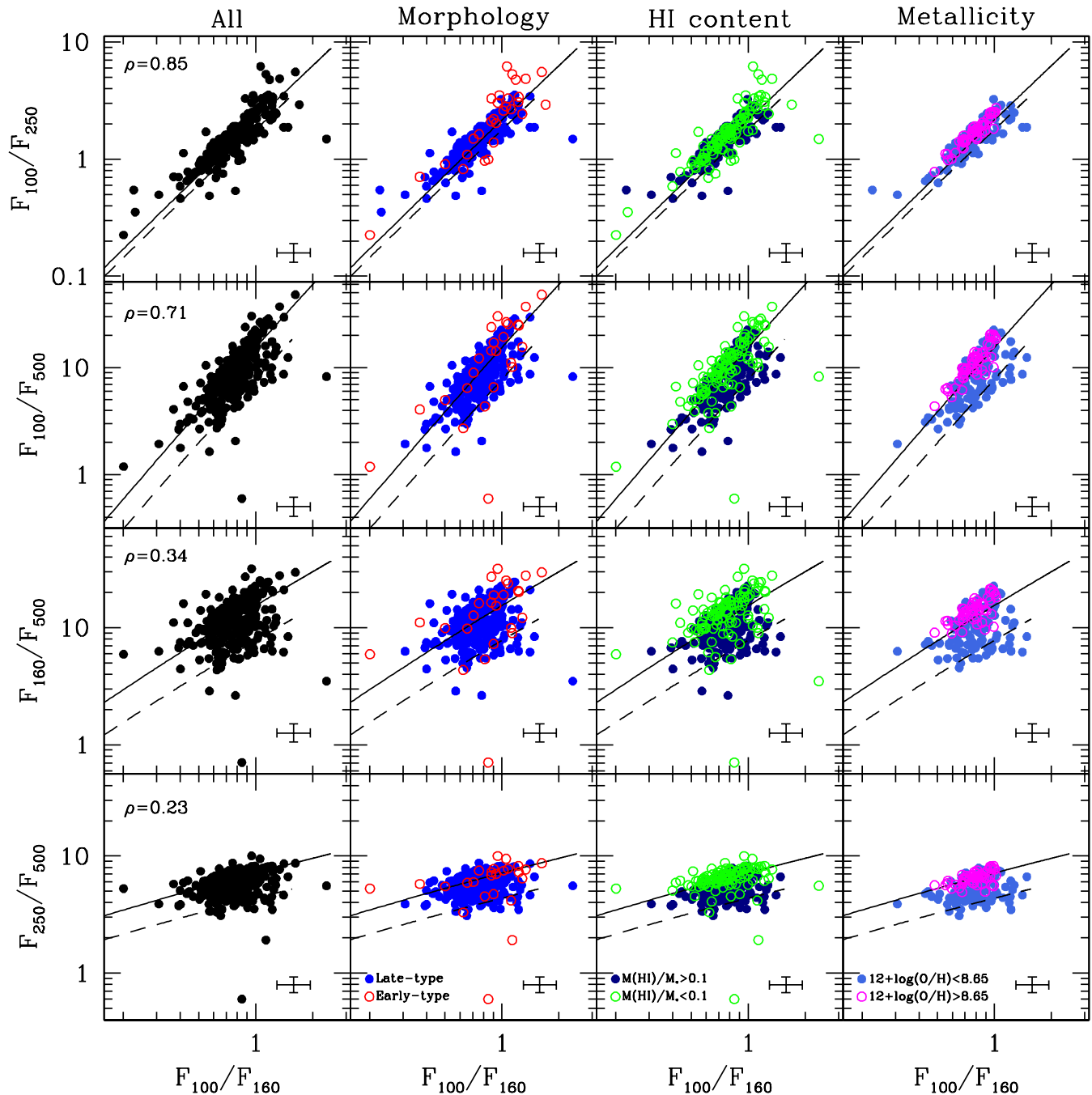}
\caption{From top to bottom: 100-to-250 $\mu$m, 100-to-500 $\mu$m, 160-to-500 $\mu$m and 250-to-500 $\mu$m as a 
function of the 100-to-160 $\mu$m flux density ratio. 
The first column shows the entire HRS sample, while in the following three columns points are colour-coded according to morphological type (open circles=E+S0, filled circles=Sa and later), 
\hi\ gas fraction (open circles=log($M(HI)/M_{*}$)$<$-1, filled circles=log($M(HI)/M_{*}$)$>$-1) and gas phase metallicity (open circles=12+log(O/H)$>$8.65, 
filled circles=12+log(O/H)$<$8.65). The Pearson correlation coefficients ($\rho$) for the whole sample are shown in the top left corner of each panel. 
The solid and dashed lines represent the expected colours for a modified black body with $\beta$=2 and 1, respectively. 
We consider a temperature range between 10 and 40 K. Typical errorbars are shown on the bottom 
right corner of each panel.}
\label{colours}
\end{figure*}
\setcounter{table}{1}
\begin{table*}
\caption {The Pearson correlation coefficients ($\rho$) and scatter ($\sigma$) of the best-fitting bisector linear fit for each sample shown in the colour-colour relations of Fig.~\ref{colours}. }
\[
\label{tablestat}
\begin{array}{lccccccccccccccc}
\hline\hline
\noalign{\smallskip}
Sample & \multicolumn{3}{c}{\rm F_{100}/F_{160} - F_{100}/F_{250}} & ~ & \multicolumn{3}{c}{\rm F_{100}/F_{160} - F_{100}/F_{500}} &~&  \multicolumn{3}{c}{\rm F_{100}/F_{160} - F_{160}/F_{500}}& ~&\multicolumn{3}{c}{\rm F_{100}/F_{160} - F_{250}/F_{500}}\\
\noalign{\smallskip}
\hline
 & N &\rho & \sigma &~& N & \rho & \sigma &~& N & \rho & \sigma &~& N & \rho & \sigma \\
\hline
All                                  & 282 & 0.84 & 0.05 &~& 274 &0.70 & 0.10 &~& 274 &0.30 & 0.14 &~& 274 &0.23 & 0.12 \\
Early-type                           & 29  & 0.88 & 0.09 &~&  25 &0.71 & 0.17 &~&  25 &0.39 & 0.23 &~&  25 &0.14 & 0.14 \\
Late-type                            & 253 & 0.83 & 0.05 &~& 249 &0.69 & 0.10 &~& 249 &0.27 & 0.14 &~& 249 &0.28 & 0.12 \\
M(HI)/M_{\odot}< 0.1                 & 106 & 0.84 & 0.04 &~& 101 &0.68 & 0.06 &~& 101 &0.26 & 0.09 &~& 101 &0.13 & 0.10 \\
M(HI)/M_{\odot}> 0.1                 & 169 & 0.84 & 0.06 &~& 167 &0.71 & 0.11 &~& 167 &0.34 & 0.15 &~& 167 &0.34 & 0.11 \\
12+log(O/H)>8.65                     & 50  & 0.95 & 0.02 &~&  50 &0.88 & 0.03 &~&  50 &0.66 & 0.05 &~&  50 &0.53 & 0.05 \\
12+log(O/H)<8.65                     & 112 & 0.86 & 0.06 &~& 110 &0.73 & 0.11 &~& 110 &0.32 & 0.15 &~& 110 &0.33 & 0.10 \\

\noalign{\smallskip}
\hline
\hline

\end{array}
\]
\end{table*}

Finally, we compared our PACS 100 $\mu$m flux density estimates with those presented in the IRAS Faint 
Source Catalogue (164 galaxies after exclusion of confused/contaminated objects), 
finding an average difference of $\sim$+7$\pm$15\% (see also \citealp{pacs_iras}). 

We remind the reader that, although the central wavelengths of MIPS and IRAS correspond to those of PACS, the bandpasses 
are not identical and part of the offsets shown above are certainly due to the different filter responses 
of the three instruments.

\section{Far-infrared/sub-millimeter colours as a proxy for the shape of the dust SED}
In the last few years, several studies have shown how infrared colours can be used as a proxy of dust properties 
(e.g., \citealp{boselli10,boselli12,dale12,bendo10,bendo12b,galametz2010,boquien11,remy13}). 
The novelty of the present work is that, for the first time, 
we cover the 100-500 $\mu$m domain for a representative 
sample of galaxies spanning a large range in stellar mass, star formation activity, cold gas and 
metal content. For example, compared to the work presented in \cite{boselli12}, which focused on 
\hi-normal spiral galaxies only, this analysis takes advantage of a more complete coverage at wavelengths 
shorter than 250 $\mu$m, and includes the entire HRS sample detected by {\it Herschel} (282 versus 146 objects). 
Similarly, the number of HRS galaxies detected at all PACS and SPIRE wavelengths is significantly 
larger (i.e., 282 versus 195) than that of \cite{auld13}, which focuses on Virgo cluster galaxies
only. 

Particularly interesting is to quantify how well the shapes of the dust SED at the short and long wavelength-ends 
correlate among each other. Indeed if, in the 100-500 $\mu$m wavelength range, the dust SED can be well approximated 
by a single modified black-body with fixed $\beta$ (i.e., the variation of the dust emissivity with frequency 
described by $\kappa_{\nu}=\kappa_{0}\times(\nu/\nu_{0})^{\beta}$), all FIR/sub-mm colours should be strongly correlated. 

The SPIRE flux densities are obtained from \cite{ciesla12}, but we applied several 
corrections to these flux estimates. We multiplied their values by 1.0253, 1.0250 and 1.0125 
at 250, 350 and 500 $\mu$m to take into account the new SPIRE calibration 
(v.11), and then by  0.9097, 0.9136 and 0.8976 at 250, 350 and 500 $\mu$m, to correct  
for the new beam areas \citep{spirecal,spireDR}.
We did not make any attempt to include variations of the beam size as a function of the 
shape of the SED \citep{spireDR}, as 
these are generally within the measurement errors (\ls 10\%). Moreover, such correction would mainly result in 
a systematic offset in the flux densities, whereas the relative variation between the SPIRE bands would be \ls 3\% for the 
ranges of $\beta$ investigated here. Thus, we are confident that this does not affect our conclusions.   
\begin{figure*}
\centering
\includegraphics[width=17.8cm]{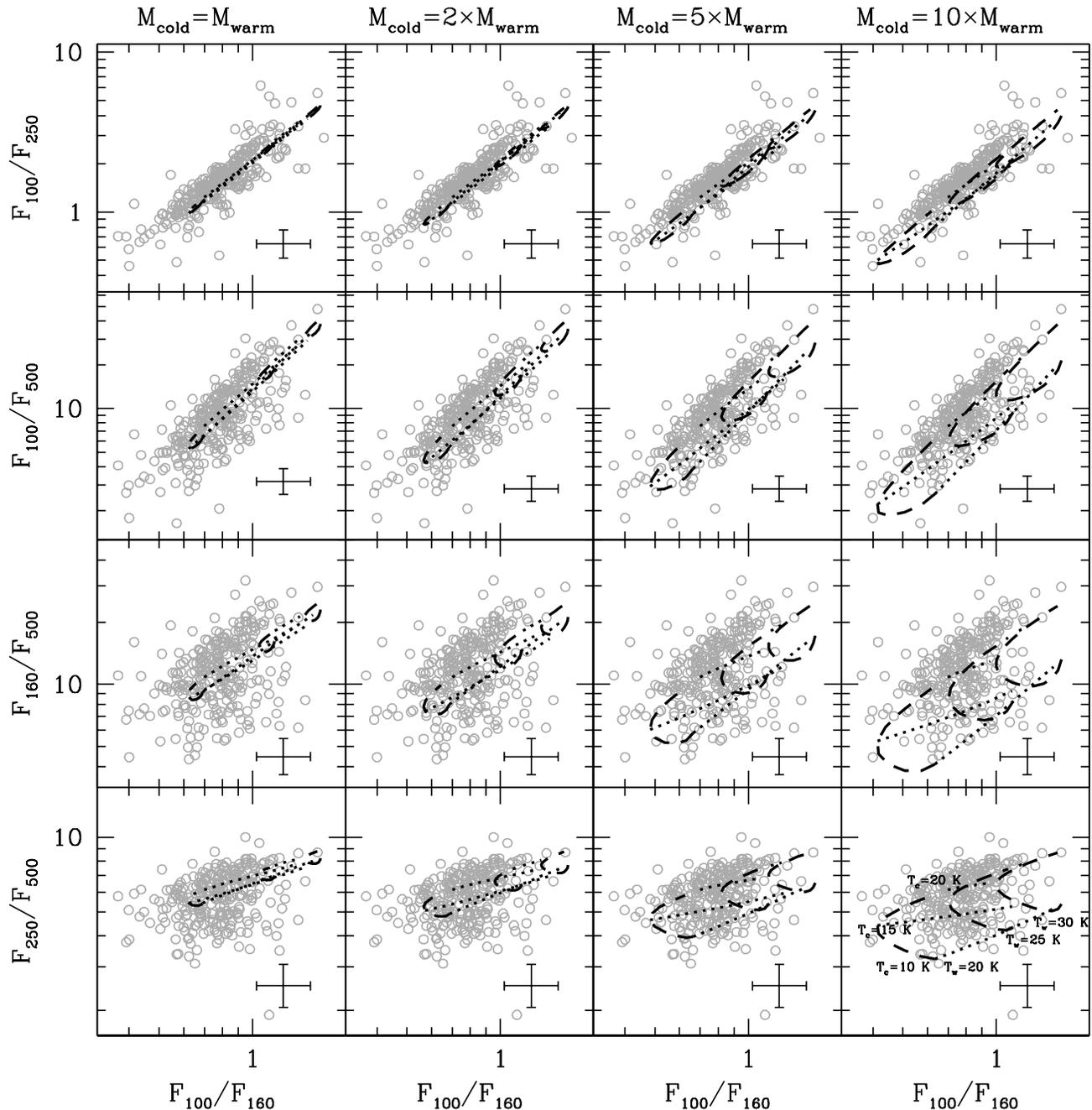}
\caption{Same as Fig.~\ref{colours}, but with the predictions for two temperatures modified black-body SEDs with
$\beta$=2 overplotted on the data points. In each plot, 
isotherms for the cold ($T_{c}$=10, 15 and 20 K) and warm ($T_{w}$=20, 25 and 30K) dust components are indicated by the dotted and dashed lines, 
respectively. Cold-to-warm dust mass ratios are 1, 2, 5 and 10 from left to right.}
\label{2BB}
\end{figure*}

In Fig.~\ref{colours} we plot the 100-to-160 $\mu$m flux density ratio, which usually embraces the peak of the dust SED, 
as a function of various flux density ratios (i.e., from top to bottom: 100-to-250 $\mu$m, 100-to-500 $\mu$m, 160-to-500 $\mu$m and 250-to-500 $\mu$m) sensible 
to the shape of the SED at increasingly longer wavelengths\footnote{In order to avoid the need to apply colour corrections when comparing 
with model predictions, here we plot the ratio of the responsivity function-weighted flux density measurements.}.
Similar results are found if additional colours (e.g., including the 350$\mu$m flux density) are used.
\begin{figure*}
\centering
\includegraphics[width=17.8cm]{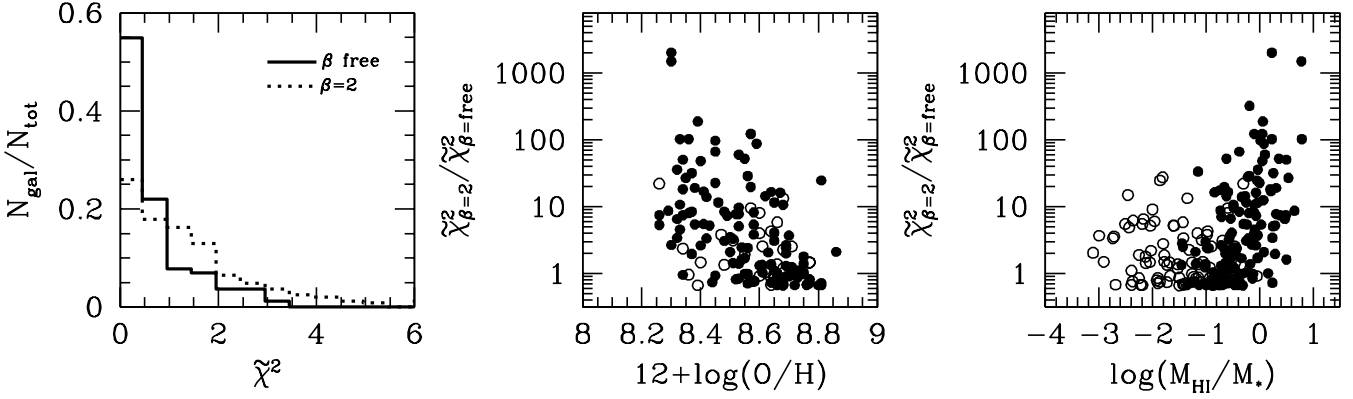}
\caption{The left panel shows the distribution of reduced $\chi^{2}$ ($\tilde{\chi^{2}}$) for the best-fitting single modified black-body with $\beta$=free (solid line) 
and $\beta$=2 (dashed). The ratio of $\tilde{\chi^{2}}$ obtained for the two cases as a function of 
gas-phase metallicity and \hi\ gas fraction are presented in the central and right panel, respectively. 
Empty circles show galaxies with \hi\ deficiency greater than 0.5. We show only those galaxies for 
which at least one of the two $\tilde{\chi^{2}}$ corresponds to a probability $P\geq$95\%.}
\label{chi2}
\end{figure*}

It is clear that the farther away in wavelength two colours are, the weaker their correlation is, 
as already noted by \cite{boselli12}. Indeed, the Pearson correlation coefficient ($\rho$) decreases from $\sim$0.8 to $\sim$0.2 when 
moving from the 100-to-250 $\mu$m to the 250-to-500 $\mu$m flux density ratios (see first column of Fig.~\ref{colours}). 
Intriguingly, the increase of a factor of $\sim$3 in scatter ($\sigma$)\footnote{In order to minimize the effect of outlier galaxies, 
the scatter is defined as the interquartile of the distribution of perpendicular distances from the best-fitting bisector linear fit for each sample.} observed when moving from the top to the third panel appears 
to be due to a population of galaxies that detaches from the main relation. 
To see if this is indeed the case, in Fig.~\ref{colours} we highlight 
galaxies according to (from left to right) their morphological type, the ratio of their atomic cold gas (\hi) to stellar 
mass content and gas-phase metallicity. \hi\ measurements have mainly been obtained from \cite{haynes2011} and \cite{springob05}, and are presented 
in \cite{hrsco}\footnote{This is an updated version of the values presented in \cite{cortese11,cortese12}, which takes advantage 
of the recently published ALFALFA flux densities \citep{haynes2011} for a considerable fraction of the HRS footprint.}. 
Stellar masses are from \cite{hrsgalex}, and gas-phase metallicities (i.e., oxygen abundances) converted into the \cite{pettini04} O3N2 base metallicity 
are taken from \cite{hughes13}. We use 12+log(O/H)=8.65 (above which the stellar  vs. mass metallicity relation starts 
flattening, \citealp{kewley08}) and $M(HI)/M_{*}=$0.1 (below which the stellar mass vs. \hi\ fraction relation is no 
longer linear, \citealp{cortese11,bothwell09}) to divide gas-rich/metal-poor from gas-poor/metal-rich galaxies.
The Pearson correlation coefficients and scatter around the best-fitting bisector linear fit 
are indicated in Table \ref{tablestat}.
   
Gas-rich/metal-poor galaxies seem to be responsible for the significant increase in scatter 
when moving from the 100-to-250 $\mu$m to the 160-to-500 $\mu$m colour-colour plots.
If we consider gas-poor/metal-rich galaxies only, the scatter in the three bottom panels of Fig.~\ref{colours} decreases by at least a factor $\sim$2. 
Indeed, performing a Kolmogorov-Smirnov test, we found that there is only a $\sim$4\% chance that the 160-to-500 $\mu$m 
colour distributions of metal-poor (12+log(O/H)$<$8.65) and metal-rich (12+log(O/H)$>$8.65) 
galaxies are drawn from the same population, as already demonstrated by \cite{boselli12}. 
We note that some galaxies do not appear in the third and fourth column of Fig.~\ref{colours}. This is because for 
some objects \hi\ and metallicity information is not available.

Our findings suggest that, in the 100-500 $\mu$m regime, the shape of the dust SED for galaxies with stellar 
mass 10$^{8}$\ls M$_{*}$/M$_{\odot}$\ls10$^{12}$ cannot be reproduced by simply varying the value of the average dust 
temperature. In other words, either $\beta$ must also vary \citep{boselli12,smith12b,remy13} or multiple temperatures components 
are required \citep{shetty09b,dunne01,boquien11,bendo12b,clemens13}. 

In order to visually illustrate this result, we plot in Fig.~\ref{colours} and \ref{2BB} the colours 
expected for these two different scenarios. In Fig.~\ref{colours} we show the flux density 
ratios derived from single modified black-bodies with temperatures ranging from 
10 and 40 K and $\beta$ values fixed to 2 (solid line) and 1 (dashed line).   
In Fig.~\ref{2BB}, we show a combination of two modified black-bodies with $\beta$=2. 
We vary the cold dust temperature ($T_{c}$) from 10 to 20 K, and the warm dust temperature ($T_{w}$) from 20 to 30 K.  
The four columns show different mass ratios $M_{cold}/M_{warm}$ increasing from 1 (left) to 10 (right). 

It is clear that, while the temperature is the main driver of the trends 
observed in each colour-colour plot, only a variation in $\beta$, or an additional temperature component, 
can explain the increasing scatter when moving from the 100-to-250 $\mu$m to 160-to-500 $\mu$m colours. 
Interestingly, the two temperature components scenario is able to reproduce 
the observed range of colours only if the warm component contributes negligibly to the total dust budget 
of the galaxy (i.e., $M_{cold}/M_{warm}$\gs 5; \citealp{Vlahakis05}). 
This is easy to understand if we consider the fact that, at fixed dust mass, 
the flux density emitted by a black-body in the FIR/submm wavelength range  
increases with temperature. Thus, if the warm and cold components have the same dust mass, 
the warm dust dominates the total emission, and the shape of the SED is very close to 
that of a single black-body. Only if the cold dust component dominates the mass budget, 
the shape of the combined SED deviates significantly from a single black-body.
 

Unfortunately, with our current data it is impossible to discriminate between a varying $\beta$ and 
a multiple temperature component scenario. Our lack of coverage below 100 $\mu$m makes it meaningless  
to perform a two temperatures fit, as the warm component is not constrained. 
Thus, in the rest of this paper we will focus on the single modified black-body 
case only, and investigate how different assumptions on $\beta$ can affect the interpretation of {\it Herschel} 
observations. A detailed comparison with the predictions of the \cite{draine07} dust models will 
be presented in a forthcoming paper (Ciesla et al., submitted.).

\section{Fitting the dust SED with a single modified black-body}
\subsection{How well do colours trace the average dust temperature?}
The results presented in the previous section show that FIR/sub-mm colours may not 
always represent a proxy for the average underlying dust temperature. 
In order to investigate this issue in more detail, 
it is interesting to quantify how the FIR/sub-mm colours correlate with the parameters obtained from a 
single modified black-body fitting. We assume either a constant value of $\beta$=2, or keep this as a free parameter. 
The model functions were convolved with the PACS and SPIRE filter response functions and fitted to the 
relative spectral responsivity function-weighted flux density measurements. Best-fit parameters and 
their 1$\sigma$ uncertainties are determined 
via a $\chi^{2}$ minimisation using the Python version of the minimisation library \texttt{MINUIT} \citep{minuit}. 
We choose $\beta$=2 simply because this seems to correctly reproduce the shape of the SED 
for massive, metal-rich spiral galaxies in the local Universe \citep{davies11,boselli12,draine13}. 
However, our results do not qualitatively change if a different (but fixed) value of $\beta$ is used.
In the rest of the paper, we consider only those objects detected in all 5 PACS/SPIRE bands, 
and for which the reduced $\chi^{2}$ ($\tilde{\chi^{2}}$) corresponds to a probability $P\geq$95\%: i.e., $\tilde{\chi^{2}}_{dof=3}<$2.6 
(203 galaxies) and $\tilde{\chi^{2}}_{dof=2}<$3 (242 galaxies) for a fixed and variable $\beta$, respectively. 
The best-fit dust masses and temperatures for these galaxies, as well as their distance, are provided in Table~\ref{tabdust}.
This guarantees that we are not contaminated by objects whose FIR/submm emission is dominated by 
synchrotron emission \citep{baes10}.

A comparison between the reduced $\chi^{2}$ obtained for the $\beta$=free and $\beta$=2 cases 
is shown in Fig.~\ref{chi2}. Not surprisingly, leaving $\beta$ free provides on average better fits. Moreover, as shown 
in the central and right panel of Fig.~\ref{chi2}, the difference between the two techniques increases when moving 
towards metal-poor/gas-rich systems. This is even more evident when \hi-deficient galaxies (i.e., $Def_{HI}>$0.5, empty points 
in Fig.~\ref{chi2}), for which the gas content is no longer a good indicator of enrichment history \citep{cortese09,hughes13}, are excluded ($\rho=$0.38 and 0.54 for all galaxies 
and \hi-normal systems only, respectively).

\begin{figure*}
\centering
\includegraphics[width=17.5cm]{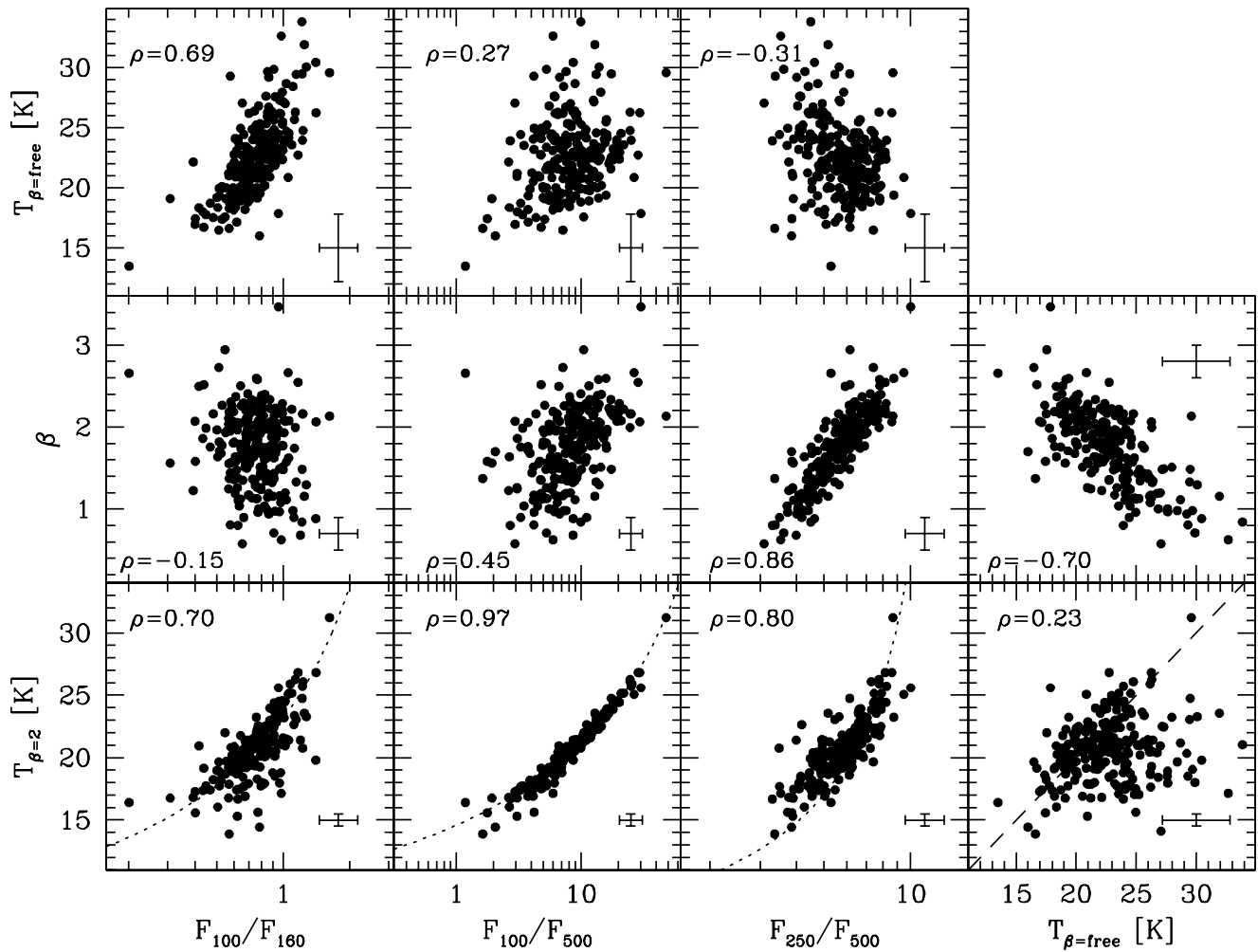}
\caption{The modified black-body best-fitting parameters as a function of far-infrared/sub-millimeter colours (from left to right: 100-to-160 $\mu$m, 100-to-250 $\mu$m, 
100-to-500 $\mu$m and 250-to-500 $\mu$m flux density ratio). The bottom row shows the dust temperature obtained by keeping $\beta$ fixed to 2, 
while the middle and top rows show the best-fitting values for $\beta$ and $T$ obtained by varying both parameters freely. 
The Pearson correlation coefficients are indicated in each panel.
In the bottom row, the dotted lines show the expected relations between temperature and colour for a single modified black-body with $\beta$=2, 
while the dashed line indicates the 1-to-1 relation.}
\label{colT}
\end{figure*}

In Fig.~\ref{colT}, we show how the FIR/sub-mm colours correlate with the best-fit parameters obtained from our SED fitting. 
Not surprisingly, all SPIRE and PACS colours strongly correlate with dust temperature if $\beta$ is kept fixed (we note that these 
results do not qualitatively change if we fix $\beta$ to a different value). It is also expected that the lowest scatter 
is observed for the colour spanning the largest wavelength range (i.e., the 100-to-500 $\mu$m flux density ratio), as 
the variation in colour is larger, and less affected by measurement errors. 

More interesting is the case when $\beta$ is treated as a free parameter. In this case, there is a clear difference in the colours 
behaviour when crossing a $\lambda$ of $\sim$200 $\mu$m. At shorter wavelengths, there is still a strong correlation of colour with temperature ($\rho\sim$0.7), 
while only a very weak trend is seen with $\beta$ ($\rho\sim$-0.15). Moving to longer wavelengths, the trends with temperature become weaker, 
and reverse for the 250-to-500 $\mu$m colour ($\rho\sim$-0.3), whereas the correlation with $\beta$ becomes gradually stronger. The best 
relation is found with the 250-to-500 $\mu$m flux density ratio ($\rho\sim$0.9), which appears to be mainly tracing variations of 
$\beta$ and not dust temperature, as also shown in Fig.~\ref{colours}. 
These results are likely a direct consequence of the fact that the FIR/submm SED for our sample peaks at $\lambda<$200 $\mu$m, and while 
the PACS colours trace the peak of the dust SED, any variations in the emissivity 
of the grains will predominantly affect the SPIRE colours. The average value of $\beta$ for HRS galaxies is $\sim$1.8$\pm$0.5, a value 
consistent with what is found in the Milky Way and in other nearby galaxies \citep{plankMW,galametz12,boselli12,smith12b,smith2013}.
\begin{figure*}
\centering
\includegraphics[width=17.5cm]{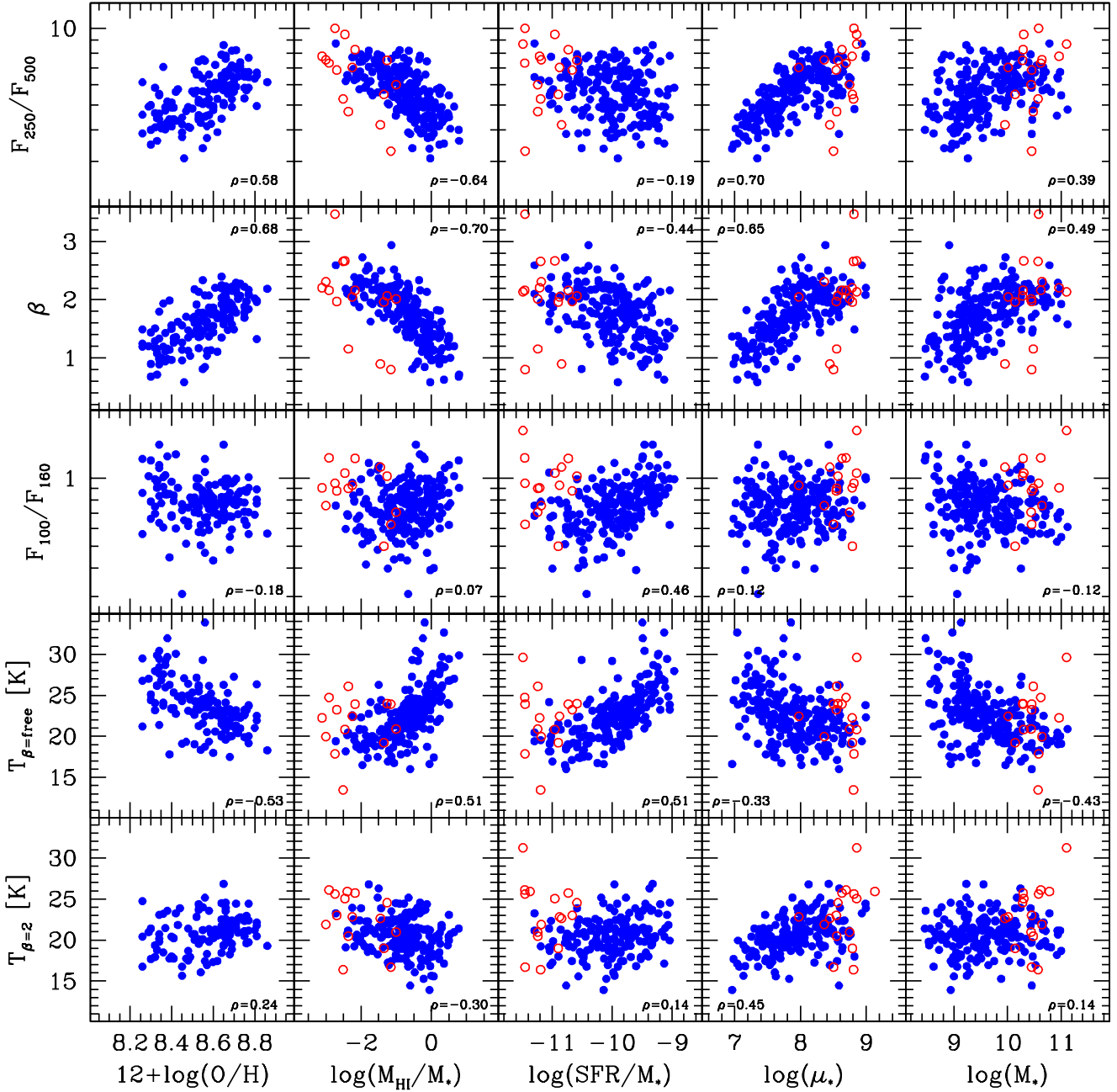}
\caption{ From top to bottom: the 250-to-500 $\mu$m flux density ratio, the best-fitting value of $\beta$, the 100-to-160 $\mu$m flux density ratio, 
the best fitting temperature assuming $\beta$=free and $\beta=$2 as a function of stellar mass, stellar mass surface density ($\mu_{*}$), 
specific star formation rate ($SFR/M_{*}$), \hi\ gas fraction ($M_{HI}/M_{*}$) and gas phase metallicity (12+log(O/H). 
Filled and open circles show late- and early-type galaxies, respectively. The Pearson correlation coefficients for the whole 
sample are shown in each panel.}
\label{ISM}
\end{figure*}

An important issue affecting any modified black-body $\chi^{2}$ fitting with $\beta$ and $T$ as free parameters is the known anti-correlation 
between them, which is clearly shown in the right column of Fig.~\ref{colT}. While it is still debated whether part 
of this anti-correlation has a physical origin \citep{shetty09b,galametz12,smith12b,juvela12,juvela13,remy13,tamatabei13}, 
there is no doubt that it is mainly due to the $\chi^{2}$ fitting technique \citep{shetty09a}. Indeed, in the 2D $\beta$ vs. $T$ plane, 
the region corresponding to the absolute minimum of $\chi^{2}$ depends on both quantities, giving 
rise to an anti-correlation between $\beta$ and $T$. This is clearly visible by just looking 
at the 2D confidence levels for any $\chi^{2}$ modified black-body fit. 
Since in the first and third columns of Fig.~\ref{colT} temperature and $\beta$ show opposite trends with colour, it is very likely that 
they are affected by this degeneracy. However, the significant difference in scatter between 
the various relations suggests that the 100-to-160 $\mu$m colour vs. $T$ and 250-to-500 $\mu$m colour vs. $\beta$ 
are less contaminated than the other correlations. As mentioned above, this is 
because the PACS colours mainly trace the peak of the dust SED, whereas the SPIRE ones are mostly 
sensitive to variations in the dust emissivity.


\subsection{The relation between dust temperature, $\beta$ and integrated galaxy properties}
In this section we investigate further how the variation of $\beta$, necessary to reproduce the observed colours of HRS galaxies 
in a single modified black-body scenario, is mirrored by a variation in galaxy properties. For comparison, we will also show 
the results obtained by keeping $\beta$ fixed, since 
we consider this an instructive exercise to illustrate how the model assumptions influence the parameters we derive. 
In Fig.~\ref{ISM}, we show how the best-fitting dust parameters, as well as the 100-to-160 $\mu$m and 250-to-500 $\mu$m flux density ratios, 
are related to gas-phase metallicities, \hi\ gas fractions, specific star formation rate ($SFR/M_{*}$), 
stellar mass surface density [$\mu_{*}$=$M_{*}/(2\pi R_{50,i}^{2})$ where $R_{50,i}$ is the radius containing 50\% 
of the total $i$-band light] and stellar mass. Star formation rates are determined by combining WISE 22$\mu$m 
(Ciesla et al., submitted.) and NUV photometry \citep{hrsgalex} using the recipes presented in \cite{hao11} as described in \cite{cortese12c}. 

By comparing the two bottom rows of Fig.~\ref{ISM}, it is clear that the assumptions made on $\beta$ significantly influence the 
correlations between temperature and integrated galaxy properties. For $\beta$ fixed to 2, the strongest correlation is found 
with stellar mass surface density ($\rho\sim$0.45). A weak anti-correlation is visible with gas-fraction ($\rho\sim$-0.3), 
while no correlation is found with specific star formation rate, stellar mass or metallicity ($\rho$ \ls0.2). 
Quite different results are obtained if $\beta$ is left free. 
In this case, the temperature anti-correlates very weakly with $\mu_{*}$ ($\rho\sim$-0.3), while it is strongly correlated with 
$SFR/M_{*}$ (see also \citealp{clemens13}), \hi\ gas fraction, metallicity and stellar mass ($\rho\sim$0.5). 
Even more importantly, some of the correlations show opposite trends. For a fixed value of $\beta$, the 
temperature increases with metallicity and stellar mass surface densities, whereas it decreases 
for $\beta$=free. The `reversal' of these correlations is driven exclusively by 
metal-poor/gas-rich galaxies, and it is simply a consequence of the fact that, for these objects,  
the best-fitting value of $\beta$ is significantly lower than 2.  
Thus, many of the correlations shown in Fig.~\ref{ISM} depend 
on the assumptions made about the dust SED, and may not be physical \citep{magnelli12,roseboom13}.


In particular, we have shown (see Fig.~\ref{colT}) that the 100-to-160 $\mu$m and 250-to-500 $\mu$m flux 
density ratios are the best proxies for $T$ and $\beta$, respectively. 
If all the trends observed in Fig.~\ref{ISM} are physical, 
we should find similar correlations when $T$ and $\beta$ are replaced by 
the flux density ratios. However, this is not always the case. 
The 100-to-160 $\mu$m flux density ratio correlates only with $SFR/M_{*}$ ($\rho\sim$0.5), 
while the 250-to-500 $\mu$m ratio correlates weakly with $SFR/M_{*}$ ($\rho\sim$-0.2), but 
varies strongly with stellar mass, stellar mass surface density, \hi\ gas fraction and gas-phase metallicity ($\rho\sim$0.6-0.7). 
Thus, the $T$ vs. \hi\ gas fraction and $\beta$ vs. $SFR/M_{*}$ trends might be spurious. 

In summary, our analysis confirms that the typical dust temperature of a 
galaxies as measured from a single modified black-body is mainly related to specific star formation rate, 
while $\beta$ varies more with the degree of metal enrichment of the ISM. 
As discussed in the previous section, at this stage it is impossible to determine whether the variation of $\beta$ across the HRS 
indicates a variation in the dust properties/composition, or it simply highlights the need of multiple temperature components 
for gas-rich/metal-poor/low-mass galaxies.

\begin{figure}
\centering
\includegraphics[width=8.5cm]{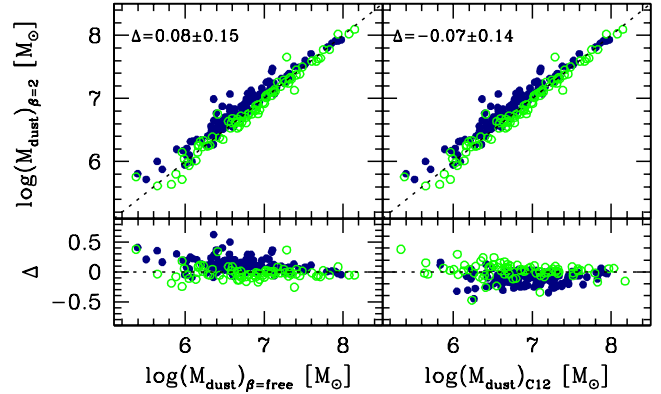}
\caption{{\it Left panel}: Comparison between the dust masses obtained from a black-body SED 
fitting with $\beta=$2 and $\beta=$free. {\it Right panel}: Dust 
masses obtained from a black-body SED fitting with $\beta=$2 as a function of those 
obtained using the empirical recipes of Cortese et al. (2012b), which are based on SPIRE colours only. 
Filled and empty circles indicate gas-rich and gas-poor galaxies, respectively (see also Fig.~\ref{colours}).}
\label{dustmass}
\end{figure}

\subsection{Dust mass estimates}
It is interesting to investigate how the variation of $\beta$ across the HRS for a single modified black-body 
affects the estimate of the dust mass reservoir. Thus, in the left panel of Fig.~\ref{dustmass}, we compare the 
dust masses obtained for $\beta$=free and $\beta$=2. Dust masses have been calculated from Eq.~\ref{eq1} assuming 
$\nu_{0}=$856.5 GHz (i.e., 350 $\mu$m) and $\kappa_{0}=$0.192 m$^{2}$ kg$^{-1}$ \citep{draine03}.
It is evident that dust masses are significantly less affected than dust temperatures by the assumptions made on $\beta$. 
The average difference between the two measurements is $\sim$0.08 dex, with a standard deviation of 0.15 dex, 
which is consistent with the typical statistical error obtained from the SED fitting: $\sim$0.05 and 0.1 dex 
for $\beta$=2 and $\beta$=free. Not surprisingly, the largest difference is observed in gas-rich galaxies 
(filled circles, $\Delta$=0.14$\pm$0.14 dex), while the two techniques give consistent results for gas-poor 
systems (empty circles, $\Delta$=-0.02$\pm$0.11 dex). 

This result implies that correlations involving 
dust masses are quite robust against the assumptions made on the shape of the SED. 
Different assumptions can certainly affect the exact slope of the dust scaling relations, 
but they are not able to produce the same dramatic inversion of some correlations observed for the dust temperature (see Fig.~\ref{ISM}).
 
This conclusion is reinforced by the fact that the differences, already quite small, between the two cases 
might be overestimated, as we varied $\beta$, by keeping fixed the value of dust opacity $\kappa_{0}$ used to determine the dust mass. 
As recently shown by \cite{bianchi13}, this is not entirely correct because the value of $\kappa_{0}$ 
is calibrated on a dust model with a well defined value of $\beta$. Thus, 
if $\beta$ changes, $\kappa_{0}$ should change as well. Unfortunately, varying $\kappa$ along with 
$\beta$ is far from trivial, and it is only possible by either having a consistent dust model for 
each value of $\beta$, or by comparing dust mass estimates obtained from SED fitting with the ones obtained from 
other independent methods: e.g., using the amount of cold gas and metals, as proposed by \cite{james02}.

Finally, it is interesting to compare the dust masses estimated by fitting a single modified black-body 
with $\beta$=2, to those obtained by using the empirical recipes developed by \cite{cortese12}, which 
assume $\beta$=2 but are based on SPIRE data only. In this way we can quantify the benefit provided 
by inclusion of the PACS data in the dust mass estimates. As shown in the right panel of Fig.~\ref{dustmass}, 
the two estimates show a good agreement with a mean difference of -0.07 dex and a standard deviation of 0.14 dex, 
lower than the typical uncertainty of 0.2 dex in the recipes by \cite{cortese12}. Even in this case, the largest 
offset (-0.12$\pm$0.11 dex) is found for gas-rich galaxies. This is a natural consequence of the 
fact that, for these objects, the shape of the dust SED is no longer perfectly consistent with $\beta$=2.

Thus, while dust mass estimates based on SPIRE colours are a reliable tool for estimating 
dust masses within 0.2 dex, only a complete coverage of the 100-500 $\mu$m wavelength range 
can provide us with accurate (within 0.1dex) dust mass estimates necessary to 
quantify in great detail the correlation between dust mass and other galaxy properties.

\section{Summary \& Conclusions}
In this paper we presented PACS 100 and 160 $\mu$m 
integrated photometry for the {\it Herschel} Reference Survey. 
We have combined these data with SPIRE observations 
to investigate how the shape of dust SED varies across the Hubble sequence.
Being the largest representative sample of nearby galaxies 
with homogeneous coverage in the 100-500 $\mu$m wavelength domain, the 
HRS is ideal to quantify if and how dust emission varies across 
the local galaxy population. 
Our main results are as follows.

\begin{itemize}

\item The shape of the dust SED 
is not well described by a single modified black-body having just 
the dust temperature as a free parameter. 
Instead, there is a clear need to vary the dependence of the dust 
emissivity ($\beta$) on wavelength, or to invoke 
multiple temperature components in order to 
reproduce the colours observed in our sample.
This is particularly important as the HRS does not include 
very metal-poor dwarf galaxies, for which we already knew 
that the dust SED is significantly different from the one of metal-rich, 
massive galaxies \citep{galliano2005,galliano11,engelbracht2008,galametz2009,remy13}.
Our results suggest that the difference in FIR/sub-mm 
colours between giant and dwarf galaxies \citep{draine07b} may not be 
the result of a dramatic transition in dust properties, 
but just the consequence of the gradual variation that 
we observe as a function of metal and gas content.

\item The variation in the slope of the dust SED strongly affects 
dust temperature estimates from single modified black-bodies fits. 
In particular, the correlations between galaxy properties and 
dust temperatures strongly depend on the assumptions made on $\beta$: i.e., 
trends can disappear or even reverse. Conversely, dust mass estimates are more robust, and 
variations in $\beta$ do not produce the same dramatic inversion 
of some correlations observed for the dust temperature.

\item We confirm that the temperature of a 
single modified black-body is mainly related to specific star formation rate, 
while $\beta$ varies more with the degree of metal enrichment of the ISM.

\end{itemize}

The results presented in this paper may appear in contradiction 
with several recent works showing that the dust SED is 
very well reproduced by a simple modified black-body with 
$\beta=$2 \citep{davies11,auld13}. However, all these works were focused 
on massive, metal-rich and relative gas-poor galaxies, for which 
we also find that a constant value of $\beta$ provides a good 
fit to our data. It is when we move to the gas-rich/metal-poor 
regime that the shape of the SED starts to change \citep{boselli10,boselli12,remy13}. 

Our findings overall reinforce the results already presented in \cite{boselli10,boselli12}. 
However, it is important to note that the discovery of a clear variation in the shape of the SED across the HRS 
has only been possible thanks to the large wavelength coverage obtained by combining both PACS and SPIRE data. 
Indeed, with SPIRE or PACS data only, it would be not only much more difficult to show under which conditions 
a simple modified black-body approach does not work, but it would also be nearly impossible to quantify 
how model assumptions can affect the correlation of 
dust temperature with star formation, galaxy structure and chemical enrichment.

\section*{Acknowledgments}
We thank an anonymous referee for his/her very useful comments and suggestions 
which have significantly improved this manuscript.
LC thanks B. Draine for useful discussions, and B. Catinella for comments on this manuscript.
We thank all the people involved in the construction and the launch of {\it Herschel}.

The research leading to these results has received funding from the European CommunityÕs Seventh Framework Programme
(/FP7/2007-2013/) under grant agreement No 229517, and was supported under Australian Research Council's Discovery 
Projects funding scheme (project number 130100664).
IDL is a postdoctoral researcher of the FWO-Vlaanderen (Belgium).

PACS has been developed by a consortium of institutes led by MPE (Germany) and including UVIE (Austria); KU Leuven, CSL, IMEC (Belgium); 
CEA, LAM (France); MPIA (Germany); INAF-IFSI/OAA/OAP/OAT, LENS, SISSA (Italy); IAC (Spain). This development has been supported 
by the funding agencies BMVIT (Austria), ESA-PRODEX (Belgium), CEA/CNES (France), DLR (Germany), ASI/INAF (Italy), 
and CICYT/MCYT (Spain).
SPIRE has been developed by a consortium of institutes led by Cardiff University (UK) and including Univ. Lethbridge (Canada); 
NAOC (China); CEA, LAM (France); IFSI, Univ. Padua (Italy); IAC (Spain); Stockholm Observatory (Sweden); Imperial College London, 
RAL, UCL-MSSL, UKATC, Univ. Sussex (UK); and Caltech, JPL, NHSC, Univ. Colorado (USA). This development has been supported by 
national funding agencies: CSA (Canada); NAOC (China); CEA, CNES, CNRS (France); ASI (Italy); MCINN (Spain); SNSB (Sweden); STFC (UK); and NASA (USA). 

Part of the HRS data was accessed through the Herschel Database in Marseille (HeDaM - \url{http://hedam.lam.fr}) 
operated by CeSAM and hosted by the Laboratoire d'Astrophysique de Marseille.

We acknowledge the use of the NASA/IPAC Extragalactic Database (NED) which is operated by the 
Jet Propulsion Laboratory, California Institute of Technology, under contract with the National 
Aeronautics and Space Administration.

\bibliography{main}

\onecolumn
\setcounter{table}{0}
\LTcapwidth=\textwidth
\begin{landscape}
\begin{center}
\tiny
\begin{longtable}{cccccccccccccccccccc}
\caption[The PACS 100 and 160$\mu$m flux densities of the HRS.]{The PACS 100 and 160$\mu$m flux densities of the HRS.\label{tab}}\\
\hline \hline \\[-2ex]
  \multicolumn{1}{c}{HRS} &
  \multicolumn{1}{c}{CGCG} &
  \multicolumn{1}{c}{VCC} &
  \multicolumn{1}{c}{UGC} &
  \multicolumn{1}{c}{NGC} &
  \multicolumn{1}{c}{IC} &
  \multicolumn{1}{c}{R.A. (J.2000)} &
  \multicolumn{1}{c}{Dec (J.2000)} &
  \multicolumn{1}{c}{Type} &
  \multicolumn{1}{c}{Flag$_{100}$} &
  \multicolumn{1}{c}{F$_{100}$} &
  \multicolumn{1}{c}{$\sigma_{100}$} &
  \multicolumn{1}{c}{Flag$_{160}$} &
  \multicolumn{1}{c}{F$_{160}$} &
  \multicolumn{1}{c}{$\sigma_{160}$} &
  \multicolumn{1}{c}{a} &
  \multicolumn{1}{c}{b} &
  \multicolumn{1}{c}{P.A.} &
  \multicolumn{1}{c}{Proposal ID} \\
  
  \multicolumn{1}{c}{} &
  \multicolumn{1}{c}{} &
  \multicolumn{1}{c}{} &
  \multicolumn{1}{c}{} &
  \multicolumn{1}{c}{} &
  \multicolumn{1}{c}{} &
  \multicolumn{1}{c}{hh:mm:ss.ss} &
  \multicolumn{1}{c}{dd:mm:ss.s} &
  \multicolumn{1}{c}{} &
  \multicolumn{1}{c}{} &
  \multicolumn{1}{c}{Jy} &
  \multicolumn{1}{c}{Jy} &
  \multicolumn{1}{c}{} &
  \multicolumn{1}{c}{Jy} &
  \multicolumn{1}{c}{Jy} &
  \multicolumn{1}{c}{arcsec} &
  \multicolumn{1}{c}{arcsec} &
  \multicolumn{1}{c}{degree} &
  \multicolumn{1}{c}{} \\[0.5ex] \hline
   \\[-1.8ex]
\endfirsthead
\multicolumn{19}{c}{{\tablename} \thetable{} -- Continued.} \\[0.5ex]
  \hline \hline \\[-2ex]
  \multicolumn{1}{c}{HRS} &
  \multicolumn{1}{c}{CGCG} &
  \multicolumn{1}{c}{VCC} &
  \multicolumn{1}{c}{UGC} &
  \multicolumn{1}{c}{NGC} &
  \multicolumn{1}{c}{IC} &
  \multicolumn{1}{c}{R.A. (J.2000)} &
  \multicolumn{1}{c}{Dec (J.2000)} &
  \multicolumn{1}{c}{Type} &
  \multicolumn{1}{c}{Flag$_{100}$} &
  \multicolumn{1}{c}{F$_{100}$} &
  \multicolumn{1}{c}{$\sigma_{100}$} &
  \multicolumn{1}{c}{Flag$_{160}$} &
  \multicolumn{1}{c}{F$_{160}$} &
  \multicolumn{1}{c}{$\sigma_{160}$} &
  \multicolumn{1}{c}{a} &
  \multicolumn{1}{c}{b} &
  \multicolumn{1}{c}{P.A.} &
  \multicolumn{1}{c}{Proposal ID} \\
  
  \multicolumn{1}{c}{} &
  \multicolumn{1}{c}{} &
  \multicolumn{1}{c}{} &
  \multicolumn{1}{c}{} &
  \multicolumn{1}{c}{} &
  \multicolumn{1}{c}{} &
  \multicolumn{1}{c}{hh:mm:ss.ss} &
  \multicolumn{1}{c}{dd:mm:ss.s} &
  \multicolumn{1}{c}{} &
  \multicolumn{1}{c}{} &
  \multicolumn{1}{c}{Jy} &
  \multicolumn{1}{c}{Jy} &
  \multicolumn{1}{c}{} &
  \multicolumn{1}{c}{Jy} &
  \multicolumn{1}{c}{Jy} &
  \multicolumn{1}{c}{arcsec} &
  \multicolumn{1}{c}{arcsec} &
  \multicolumn{1}{c}{degree} &
  \multicolumn{1}{c}{} \\[0.5ex] \hline
   \\[-1.8ex]
\endhead 
 \\[-1.8ex] \hline
  \multicolumn{19}{c}{{Continued on the next page\ldots}} \\
\endfoot

  \\[-1.8ex] \hline \hline
\endlastfoot

  1 & 123-035 & 0 & 0 & 0 & 0 & 10:17:39.66 & +22:48:35.9 & 13 & 1 & 0.748 & 0.169 & 1 & 0.932 & 0.079 & 48. & 40. & -10. & OT1\_lcortese\_1\\
  2 & 124-004 & 0 & 5588 & 0 & 0 & 10:20:57.13 & +25:21:53.4 & 5 & 1 & 2.439 & 0.227 & 1 & 2.808 & 0.179 & 47. & 45. & 40. & OT1\_lcortese\_1\\
  3 & 94-026 & 0 & 5617 & 3226 & 0 & 10:23:27.01 & +19:53:54.7 & 0 & 0 & 0.0 & 0.0 & 2 & 0.846 & 0.087 & 39. & 39. & 15. & GT1\_lspinogl\_2\\
  4 & 94-028 & 0 & 5620 & 3227 & 0 & 10:23:30.58 & +19:51:54.2 & 3 & 2 & 17.589 & 1.104 & 2 & 22.675 & 1.165 & 100. & 84. & -25. & GT1\_lspinogl\_2/OT2\_aalonsoh\_2\\
  5 & 94-052 & 0 & 0 & 0 & 610 & 10:26:28.37 & +20:13:41.5 & 7 & 1 & 4.502 & 0.331 & 1 & 5.563 & 0.528 & 81. & 48. & 28. & OT1\_lcortese\_1\\
  6 & 154-016 & 0 & 5662 & 0 & 0 & 10:27:01.16 & +28:38:21.9 & 5 & 1 & 0.275 & 0.14 & 1 & 0.483 & 0.085 & 62. & 22. & -30. & OT1\_lcortese\_1\\
  7 & 154-017 & 0 & 5663 & 3245 & 0 & 10:27:18.39 & +28:30:26.6 & 1 & 1 & 3.472 & 0.206 & 1 & 2.843 & 0.151 & 48. & 30. & -3. & OT1\_lcortese\_1\\
  8 & 154-020 & 0 & 5685 & 3254 & 0 & 10:29:19.92 & +29:29:29.2 & 6 & 1 & 2.878 & 1.092 & 1 & 4.641 & 1.041 & 210. & 66. & 46. & OT1\_lcortese\_1\\
\end{longtable}
\end{center}
\end{landscape}

\setcounter{table}{2}
\begin{center}
\small
\begin{longtable}{lccccccc}
\caption[Best-fitting dust temperatures and masses for a single modified black-body with $\beta$=2 and $\beta$=free.]{Best-fitting dust temperatures and masses for a single modified black-body with $\beta$=2 and $\beta$=free.Only galaxies for which the reduced $\chi^{2}$ corresponds to a probability $P\geq$95\% are shown. \label{tabdust}}\\
\hline \hline \\[-2ex]
  \multicolumn{1}{c}{HRS} &
  \multicolumn{1}{c}{D} &
  \multicolumn{2}{c}{$\beta$=2} &
  \multicolumn{1}{c}{} &
  \multicolumn{3}{c}{$\beta$=free} \\

  \multicolumn{1}{c}{} &
  \multicolumn{1}{c}{} &
  \multicolumn{1}{c}{T} &
  \multicolumn{1}{c}{log(M$_{dust}$)} &
  \multicolumn{1}{c}{} &
  \multicolumn{1}{c}{$\beta$} &
  \multicolumn{1}{c}{T} &
  \multicolumn{1}{c}{log(M$_{dust}$)} \\

  \multicolumn{1}{c}{} &
  \multicolumn{1}{c}{Mpc} &
  \multicolumn{1}{c}{K} &
  \multicolumn{1}{c}{M$_{\odot}$} &
  \multicolumn{1}{c}{} &
  \multicolumn{1}{c}{} &
  \multicolumn{1}{c}{K} &
  \multicolumn{1}{c}{M$_{\odot}$} \\[0.5ex] \hline
   \\[-1.8ex]
\endfirsthead

\multicolumn{8}{c}{{\tablename} \thetable{} -- Continued.} \\[0.5ex]
  \hline \hline \\[-2ex]
  \multicolumn{1}{c}{HRS} &
  \multicolumn{1}{c}{D} &
  \multicolumn{2}{c}{$\beta$=2} &
  \multicolumn{1}{c}{} &
  \multicolumn{3}{c}{$\beta$=free} \\

  \multicolumn{1}{c}{} &
  \multicolumn{1}{c}{} &
  \multicolumn{1}{c}{T} &
  \multicolumn{1}{c}{log(M$_{dust}$)} &
  \multicolumn{1}{c}{} &
  \multicolumn{1}{c}{$\beta$} &
  \multicolumn{1}{c}{T} &
  \multicolumn{1}{c}{log(M$_{dust}$)} \\

  \multicolumn{1}{c}{} &
  \multicolumn{1}{c}{Mpc} &
  \multicolumn{1}{c}{K} &
  \multicolumn{1}{c}{M$_{\odot}$} &
  \multicolumn{1}{c}{} &
  \multicolumn{1}{c}{} &
  \multicolumn{1}{c}{K} &
  \multicolumn{1}{c}{M$_{\odot}$} \\[0.5ex] \hline
   \\[-1.8ex]
\endhead 
 \\[-1.8ex] \hline
  \multicolumn{8}{c}{{Continued on the next page\ldots}} \\
\endfoot

  \\[-1.8ex] \hline \hline
\endlastfoot
     1 &   16.79 &    21.7$^{+1.0}_{-1.0}$ &    5.93$^{+0.07}_{-0.07}$ &~&     2.0$^{+0.5}_{-0.5}$ &    21.5$^{+3.8}_{-3.1}$ &    5.94$^{+0.15}_{-0.14}$     \\
     2 &   18.44 &    24.0$^{+0.6}_{-0.6}$ &    6.27$^{+0.04}_{-0.04}$ &~&     2.3$^{+0.3}_{-0.3}$ &    21.6$^{+1.9}_{-1.6}$ &    6.36$^{+0.07}_{-0.07}$     \\
     5 &   16.71 &    22.4$^{+0.4}_{-0.4}$ &    6.62$^{+0.02}_{-0.02}$ &~&     2.2$^{+0.2}_{-0.2}$ &    20.9$^{+1.0}_{-1.0}$ &    6.68$^{+0.05}_{-0.04}$     \\
     6 &   18.89 &    13.9$^{+0.8}_{-0.8}$ &    6.80$^{+0.11}_{-0.11}$ &~&     1.4$^{+1.0}_{-0.8}$ &    16.6$^{+5.8}_{-4.0}$ &    6.57$^{+0.37}_{-0.33}$     \\
     7 &   18.77 &    25.7$^{+0.3}_{-0.3}$ &    6.26$^{+0.01}_{-0.01}$ &~&     2.2$^{+0.1}_{-0.1}$ &    23.9$^{+0.8}_{-0.8}$ &    6.31$^{+0.03}_{-0.03}$     \\
     8 &   19.37 &    15.3$^{+0.8}_{-0.7}$ &    7.48$^{+0.06}_{-0.06}$ &~&     1.3$^{+0.4}_{-0.3}$ &    20.9$^{+3.6}_{-3.2}$ &    7.11$^{+0.18}_{-0.16}$     \\
     9 &   20.21 &    21.8$^{+0.9}_{-0.9}$ &    6.53$^{+0.06}_{-0.05}$ &~&     2.5$^{+0.4}_{-0.4}$ &    18.9$^{+2.4}_{-2.0}$ &    6.66$^{+0.12}_{-0.12}$     \\
    11 &   18.93 &     -                   &    -                      &~&     1.6$^{+0.1}_{-0.1}$ &    23.7$^{+1.0}_{-1.0}$ &    6.82$^{+0.04}_{-0.04}$     \\
\end{longtable}
\end{center}

\end{document}